\documentclass[iop]{emulateapj}
\usepackage{epstopdf}
\usepackage{amsmath, amsthm, amssymb}
\usepackage{fancybox}
\usepackage{graphicx}
\usepackage{color}
\usepackage{epsfig}

\newcommand{\gr}{{\rm GR}}
\newcommand{\om}{\omega}
\newcommand{\ep}{\epsilon}
\newcommand{\per}{{\rm per}}

\newcommand{\AU}{\rm AU}
\newcommand{\af}{a_{\rm f}}

\newcommand{\ra}{\rightarrow}
\newcommand{\days}{{\,\rm d}}

\begin{document}
\title{Warm Jupiters Need Close ``Friends'' for High-Eccentricity Migration
\newline
-- A Stringent Upper Limit on the Perturber's Separation}
\author{Subo Dong\altaffilmark{1,2}, Boaz Katz\altaffilmark{2,3,4}, and Aristotle Socrates\altaffilmark{2,3}}
\altaffiltext{1}{Kavli Institute for Astronomy and Astrophysics, Peking University, Yi He Yuan Road 5, Hai Dian District, Beijing 100871, China}
\altaffiltext{2}{Institute for Advanced Study, 1 Einstein Dr.,
Princeton, NJ 08540, USA}
\altaffiltext{3}{John N. Bahcall Fellow}
\altaffiltext{4}{Einstein Fellow}

\begin{abstract}
We propose a stringent observational test on the formation of warm 
Jupiters (gas-giant planets with $10\days \lesssim P \lesssim 
100\days$) by high-eccentricity (high-$e$) migration mechanisms. 
Unlike hot Jupiters, the majority of observed warm Jupiters have 
pericenter distances too large to allow efficient tidal dissipation
to induce migration. To access the close pericenter required for  
migration during a Kozai-Lidov cycle,
they must be accompanied by a strong enough perturber to overcome 
the precession caused by General Relativity (GR), placing a strong upper limit 
on the perturber's separation. For a warm Jupiter at $a \sim 0.2\AU$, 
a Jupiter-mass (solar-mass) perturber is required to be $\lesssim3 \AU$ 
$(\lesssim 30 \AU)$ and can be identified observationally. 
Among warm Jupiters detected by Radial Velocities (RV), $\gtrsim 50\%$  {(5 out of 9)}
 with large eccentricities ($e\gtrsim 0.4$) have known Jovian 
companions satisfying this necessary condition for high-$e$ migration. In contrast, $\lesssim 20\%$ 
{(3 out of 17)} of the 
low-$e$ ($e\lesssim 0.2$) warm Jupiters have detected additional Jovian companions, 
suggesting that high-$e$ migration with planetary perturbers 
may not be the dominant formation channel.  
Complete, long-term RV follow-ups of the warm-Jupiter population
will allow a firm upper limit to be put on the fraction of these 
planets formed by high-$e$ migration. 
Transiting warm Jupiters showing spin-orbit misalignments
will be interesting to apply our test. If the misalignments 
are solely due to high-$e$ migration as 
commonly suggested, we expect that the majority of warm Jupiters with low-$e$ 
($e\lesssim0.2$) are not misaligned, in
contrast with low-$e$ hot Jupiters. 
\end{abstract}
\section{Introduction}
The origin of warm Jupiters (gas giants with period 
$10\days<P<100\days$) presents a similar puzzle to that of hot 
Jupiters ($P\lesssim 10\days$) -- neither populations can form 
{\it in-situ} according 
to popular theories of planet formation -- yet much 
less attention has been paid to the former. 

Rossiter-Mclaughlin measurements 
reveal that 
a considerable fraction of transiting hot Jupiters have orbits misaligned 
with host star spin axes \citep[e.g.][]{spin1,spin2}, which provide 
indirect support to high-eccentricity migration 
mechanisms \citep{fordrasio, wu03, ft07, secularchaos, 
supere}. These 
high-$e$ mechanisms involve the initial excitation of hot 
Jupiter progenitors at a few $\AU$ to very high eccentricity
due to gravitational perturbations by additional objects in 
the system. The excitation is then followed by successive close 
pericenter passages ($r_p \lesssim 0.05 \AU$) that drain the 
orbital energy via tidal dissipation. The hot Jupiter 
progenitors eventually become hot 
Jupiters at $a<0.1 \AU$. 

The majority of known warm Jupiters are sufficiently distant
from their hosts ($\af = a(1-e^2)>0.1\AU$) to forbid efficient tidal dissipation, 
due to the strong distance dependence of tidal effects.
However, if the orbital eccentricity of a warm Jupiter is 
experiencing Kozai-Lidov 
oscillations due to an external perturber \citep{holman, takeda05}, 
then it may be presently at the low-$e$ stage in the cycles and
over a secular timescale, reach an eccentricity high enough 
for tidal dissipation to cause significant migration 
\citep[e.g.,][]{secularchaos}. A schematic illustration
of such a high-$e$ migration scenario is shown in Fig 1 (red solid line).
{Warm Jupiters detected  by RV are shown in dots in Fig 1. within 
the black dashed lines. We define Jovian planets
to have minimum mass $M_p\sin i > 0.3 M_{\rm Jup}$ and set an 
upper limit in semi-major axis of $0.5\,\AU$ for warm Jupiters. This 
upper bound is well below the theoretical ``snow line'' of {\it in-situ} core-accretion formation at about $2.5-3\AU$ for solar-type stars 
(e.g., \citealt{snowline}) and the 
observed ``jump'' in the $a$ distribution of giant planets 
at $\sim 1\AU$ (e.g., \citealt{wright09}). In planet-planet
scattering, a Jupiter  
can migrate without tidal dissipation by factor of $\sim 2$ if another Jupiter is ejected \citep{fordrasio}, and our upper bound in distance is set to disfavor such a process.} 

We discuss a stringent observational constraint 
on warm Jupiter formation via high-$e$ migration -- 
they must be accompanied by close, easily observable perturbers. 
These close pertubers are strong enough to overcome the precession 
caused by General Relativity (GR) to reach close enough periapses for 
effective tidal dissipation within Kozai-Lidov cycles. 
In contrast, high-e migration for hot Jupiters does not subject to such a stringent constraint on perturbers. Hot Jupiter progenitors can be excited to close periapses at their initial, relatively large semi-major axes with distant perturbers, and throughout 
the subsequent migration, their periapses may keep close enough for tidal dissipation. 
Hot Jupiters formed by high-$e$ migration can thus have distant perturbers that are difficult to detect.

\section{Perturber Constraints on Warm-Jupiter High-$e$ Migration}
We derive below a lower limit on the perturber strength for 
warm-Jupiter in high-$e$ migration due to 
tidal dissipation. 
We adopt a conservative
criterion that tidal dissipation may operate when a Jovian 
planet reaches $\af = a(1-e^2) < 0.1 \AU$. Observationally, 
the eccentricities of Jovian planets circularize at $\af \sim 0.06 \AU$ 
\citep[e.g.,][]{otherQ}. Given that tidal dissipation 
has strong dependence on planet-star separation, it is 
safe to assume that tidal dissipation ceases to be efficient
when $\af > a_{\rm f, crit} = 0.1 \AU$. {We stress that the criterion presented below is a necessary but not an adequate  
condition for high-$e$ migration. 
Without satisfying the criterion, the migration cannot occur, 
while fulfilling this requirement does not guarantee migration.}

Consider a warm Jupiter with mass $M_p$ at semi-major axis $a$ and 
eccentricity $e_0$ orbiting a star with mass $M$ accompanied 
by a perturber of mass $M_{\per}$ at $a_{\per}$
and $e_{\per}$. The criterion is to require the warm Jupiter to 
reach $a(1-e^2) < a_{\rm f, crit} = 0.1\AU$ during Kozai-Lidov oscillation
 (see Fig. 2 for an example).
At a given $a$, the amplitude of Kozai-Lidov oscillation in 
eccentricity is limited by sources of precession other than those  
induced by the perturber and is insensitive to tidal dissipation. 
At $\af \sim 0.1\AU$, the precessions 
due to tides and the rotating bulge of the host are negligible 
compared to GR for 
typical hosts. Below we consider the Kozai-Lidov oscillation at the 
warm Jupiter's current $a$ due to the gravitational 
perturbation and GR precession. We ignore tidal dissipation and precession.

An analytical constraint is derived under the simplest 
assumptions: (1) quadrupole approximation in 
perturbing potential (2) the warm Jupiter treated as test particle (3) 
the equation of motion is averaged over outer and inner orbits
(``double-averaging''). We show below with numerical simulations
that these are excellent approximations in deriving this constraint.
Under these approximations, the following is a constant: \citep[e.g.,][]{ft07},
\begin{equation}\label{eq:C_K_gr}
e^2(2-5\sin^2i\sin^2\om)+\frac{\ep_{\gr}}{\sqrt{1-e^2}}={\rm const},
\end{equation}
where
\begin{align}\label{eq:ep_gr}
\ep_{\gr} &   = \frac{8GM^2b_{\per}^3}{c^2a^4M_{\per}} \cr
              &\approx 1.3 \big( \frac{M}{M_\odot} \big)^2 
              \big( \frac{a}{0.2 \AU} \big)^{-4}
              \big( \frac{M_{\per}}{M_\odot} \big)^{-1}
              \big( \frac{b_{\per}}{30 \AU} \big)^{3} \cr
              &\approx 1.4 \big( \frac{M}{M_\odot} \big)^2 
              \big( \frac{a}{0.2 \AU} \big)^{-4}
              \big( \frac{M_{\per}}{M_{\rm Jup}} \big)^{-1}
              \big( \frac{b_{\per}}{3 \AU} \big)^{3}
\end{align}
represents the relative strength of GR compared to the perturber, 
$i$ is the planet-pertuber mutual inclination, $\omega$ is planet's 
argument of pericenter, and $b_{\per} = a_{\per}(1-e_{\per}^2)^{1/2}$ is perturber's 
semi-minor axis. 

From Eq.~(\ref{eq:C_K_gr}), to reach an 
eccentricity $e$ from $e_0$, the following criterion must be 
satisfied regardless of the values of $i$ and $\omega$,
\begin{equation}\label{eq:eminmax}
\ep_{\gr}
\left(
\frac{1}{\sqrt{1-e^2}}-
\frac{1}{\sqrt{1-e_0^2}}
\right)
< 2e_0^2 + 3e^2.
\end{equation}

We then put a lower limit on the ``strength'' of the 
perturber to reach $a(1-e^2) < a_{\rm f,crit}$ (and an upper limit
on the separation ratio between the perturber and the warm 
Jupiter), 
\begin{align}
\label{eq:crit}
&\frac{b_\per}{a}< 
\left(\frac{8GM}{c^2a}\right)^{-1/3}
\left(\frac{M}{M_\per}\right)^{-1/3}\times\cr
&\left[2e_0^2+3\left(1-\frac{a_{\rm f,crit}}{a}\right)\right]^{1/3}
\left(\sqrt{\frac{a}{a_{\rm f,crit}}}-\frac{1}{\sqrt{1-e_0^2}}\right)^{-1/3}.
\end{align}
Fig.3 shows the constraints on $b_{\per}$
for $M_{\per} = M_\odot$ and $M_{\per} = M_{\rm Jup}$ 
in the upper and lower panels respectively,
derived from Eq.~\ref{eq:crit} for $a_{\rm f,crit} = 0.1\AU$.
The blue lines from above to below are for $e_0 = 0.5, 0.3, 0.0$, 
respectively ($e_0=0.3$ in dashed lines while others in solid lines).

Recently, it was realized that corrections due to various 
approximations above may lead to significant effects in 
several scenarios \citep[e.g.][]{ford00, naoznature, katzoct, 
lithoct, katzdong}.

We show that the analytic constraint given by Eq. 4 are not 
affected by the inaccuracies of the adopted approximations
 using numerical integrations without these approximations. 
 The effects of the quadrupole and test particle approximations 
 are studied by performing 
20000 simulations (10000 for $M_\per =M_{\odot}$ and 10000 for 
$M_\per=M_{\rm Jup}$). These simulations employ the double averaged approximation 
but include the octupole term and are not restricted to the test 
particle approximation. The warm Jupiters have $e_0=0.3$ and $a$ uniformly 
distributed (randomly) between $0.15$ and $0.5$AU. The eccentricities of the outer 
perturbers are uniformly distributed within $0-0.5$. The ratio $b_\per/a$ are uniformly 
distributed within $100-300$ ($10-30$) for solar-mass  
(Jupiter-mass) perturbers. The orbital orientations of the outer and 
inner orbits are randomly distributed isotropically. All runs 
were integrated to $5\,\rm Gyrs$. The results are 
shown in Fig 3. The integrations in which the warm Jupiter 
reaches $a(1-e^2)<a_{\rm f,crit}=0.1\rm AU$ are plotted as red 
dots and others in black. The analytical constraint given 
by Eq.~\ref{eq:crit} for the appropriate eccentricity $e_0=0.3$ (dashed blue lines) 
accurately traces the border of required perturbers for achieving 
the required eccentricity. {Note that at the limit given in Eq.\ref{eq:crit}, 
the strength of the octupole 
, $\sim \ep_{\rm oct} = a/a_\per[e_{\per}/(1-e_{\per}^2)] \lesssim 
1/10 (M_\per/M_{\rm Jup})^{-1/3}$, is negligible compared to 
GR, $\ep_{\gr} \sim 1$. While a small octupole can change the orbital inclination and lead to Kozai cycles with growing eccentricities, the eccentricity cannot surpass the limit Eq. \eqref{eq:crit}, which is the maximal value for all mutual orientations.} \footnote{
See relevant discussion in 
``Maximal $e$ and General Relativity (GR) precession'' of Katz, Dong \& Malhotra., 2011, arXiv:1106.3340.}

For the scenarios considered, the Kozai-Lidov time scale is 
much longer than the outer (and inner) orbital time scales, 
justifying the double averaging assumption. To illustrate 
this, the results of a direct 3-body integration are compared to those of 
an double-averaging integration in Fig 2. 
The considered warm Jupiter is at $a=0.3\rm AU$ and $e_0=0.3$ and has 
a solar-mass perturber corresponding to the limit derived from 
Eq.~\ref{eq:crit} with $e_\per=0.5$ and $b_\per=68.8 \rm AU$. The initial 
inclination is at $90^\circ$. The results of two 
integrations are practically indistinguishable (black dashed: double-averaging;
red solid: direct 3-body), validating 
the double-averaging approximation. The approximation 
is even better for a Jupiter-mass perturber satisfying the same constraint. 
This is because it has a shorter period and a similar Kozai-Lidov time scale.

In the limit of $a\gg a_{\rm f,crit}$ and $e_0\ra0$, the following useful approximation can be obtained using Eq.~\ref{eq:crit},
\begin{align}
\label{eq:minper}
\frac{b_\per}{a}
&<
\left(\frac{8GM}{3c^2 \sqrt{a a_{\rm f,crit}}}\right)^{-1/3}
\left(\frac{M}{M_\per}\right)^{-1/3} \cr
&\approx 175 
\left(\frac{M_\per}{M_\odot}\right)^{1/3}
\left(\frac{M}{M_\odot}\right)^{-2/3}
\left(\frac{a}{0.2\AU}\right)^{1/6}
\left(\frac{a_{\rm f,crit}}{0.1\AU}\right)^{1/6}\cr
&\approx 17 
\left(\frac{M_\per}{M_{\rm Jup}}\right)^{1/3}
\left(\frac{M}{M_\odot}\right)^{-2/3}
\left(\frac{a}{0.2\AU}\right)^{1/6}
\left(\frac{a_{\rm f,crit}}{0.1\AU}\right)^{1/6}.
\end{align}
{We stress that this approximation should be used for order-of-magnitude estimates since it is 
only accurate in the limit $e\sim 0$ and $a\gg a_{\rm f,crit} = 0.1 \AU$.}

For warm Jupiters with $a\sim 0.1\AU-0.5\AU$ with Jovian-planet perturbers, 
this constraint leads to an 
upper limit in orbital separation of $\sim 1.5-10 \AU$ 
(period $2-30 {\rm yr}$). The RV semi-amplitude is 
$\gtrsim 10 {\rm m/s}$, accessible to available high-precision 
RV instruments. The perturbers at the high end in period range
($\sim 20 - 30 {\rm yr}$) are more challenging since they may not 
have completed the full orbits yet during the monitoring 
projects. While for more massive perturbers, the upper limit
in orbital separation implies much longer periods -- 
($P_{\per} \propto {M_\per}^{1/2}$), they can generally be 
identified from the easily detectable RV linear trends,
\begin{align}
|\dot{v_\perp}|
&= |\frac{GM_\per}{r_\per^2} \sin i_\per \sin \theta_\per|\\
&\sim 200 {\,\rm m s^{-1} yr^{-1}} \big( \frac{M_{\per}}{1 M_\odot} \big)
\big( \frac{r_{\per}}{30 \AU} \big)^{-2} \nonumber\\
&\sim 20 {\,\rm m s^{-1} yr^{-1}} \big( \frac{M_{\per}}{1 M_J} \big)
\big( \frac{r_{\per}}{3 \AU} \big)^{-2} \nonumber
\end{align}
where $i_\per$, $\theta_\per$ and $r_\per$ are the inclination,
position angle with respect to the line of the node and
orbital separation of the perturber, respectively. 

\section{Observations \& Discussion}
In RV surveys, close binaries are commonly excluded to avoid
contamination of the spectra, making the bias for estimating
stellar perturbers challenging. We focus on planetary perturbers.
There are $34$ warm Jupiters discovered with RV at $a_{\rm f}>0.1\AU$ 
and $a<0.5\AU$ listed in the exoplanets.org database \citep{exoplanetsorg}.  
$10$ of them have additional Jovian planets at longer orbits (see Fig.1, 
in which planets with outer Jovian companions 
are plotted in cyan). All perturbers satisfy the 
constraint in Eq.~\ref{eq:crit}.  Fig.4 shows the eccentricity
distribution for all warm Jupiters (blue dashed) and for those with 
external Jovian perturbers (red solid). 
The fraction of warm Jupiters with detected
Jovian perturbers appears to be a growing function of their 
eccentricities. This trend appears to be significant in 
spite of the uncertainties due to the small number statistics. If true, there are two 
interesting implications: (1) There seems to exist a connection 
between eccentricity and the existence of a planet perturber
capable of exciting such eccentricity. This implies that 
the eccentricities for the eccentric warm Jupiters are likely excited by 
their perturbers. High-$e$ migration is therefore an attractive 
scenario for their formation. (2) The majority of low-$e$ ($e<0.2$) 
warm Jupiters lack strong perturbers necessary for high-$e$ 
migration (Eq.~\ref{eq:crit}). Given that for eccentric
warm Jupiters, similar perturbers are indeed detected
around a considerable fraction of the systems, this 
deficiency seems to be unlikely due to detection 
sensitivity. Moreover, out of the five warm Jupiter systems with 
$e<0.4$ with Jovian perturbers, three are in compact multiple planet
systems with 3 or more planets (55 Cnc b, GJ 876 c and HIP 57274 c), 
which are challenging
to explain with high-$e$ migration. In contrast, for 
the five eccentric warm Jupiters at $e>0.4$, there are no 
known additional planets in the system other than their Jovian perturbers, all of which
are located further than $2\,\AU$ yet close enough to satisfy the 
constraint from Eq.\ref{eq:crit}. This is indicative that the majority   
of low-$e$ warm Jupiters are unlikely due to high-$e$ migration 
induced by planet perturbers. It is worth noting that $M_p\sin i$ rather than 
$M_p$ is constrained from RV, so the above results are statistical. 
Note that the perturbers for all 5 eccentric warm Jupiters have 
larger $M_p\sin i$ than the inner planets, consistent with simple expectations from the 
planet-planet scattering scenario that less massive planets are easier to get 
excited into high-$e$ orbits.

{We caution that the conclusions may be affected if the chance for detecting an outer perturber strongly depends on the eccentricity of the inner planet. 
The observing strategies in RV surveys can be complicated, 
especially for those involving multiple planets. For example, \citet{wright09} pointed 
out that a system was observed more frequently after a planet was found, so the detection  
of a massive planet would likely facilitate the discovery of smaller planets. A similar
selection effect may make the detection of perturbers of eccentric warm Jupiters 
easier if their eccentricities attract particular attentions. A comprehensive sensitivity study would be helpful.  Additionally, there are a number of possible modeling degeneracies that may masquerade a 
double low-$e$ planet system as an eccentric warm Jupiter \citep{eccentric1, eccentric2}. 
Systematic modeling efforts are possibly needed to evaluate such degeneracies. There might be other mechanisms that produce
eccentric warm Jupiters associated with a perturber, including scattering
followed by disk migration similar to \citet{disk} and scattering of 3  planets with the third planet being ejected  (Petrovich et al., in prep).}

{High-precision
RV surveys $(\lesssim 5{\rm m/s})$ on thousands of stars have 
lasted for $\sim 15$ years \citep[e.g.][]{mayor11, wright09, aat}. For a 
considerable fraction of their targets, they 
can detect Jupiters at $\lesssim 5\AU$ with full orbits ({though note that some 
discoveries from these surveys remain unpublished}). Given our constraint on axis ratio of $\sim 20$ 
for Jovian perturbers, this implies the present observational constraint
on planet perturbers are likely relatively incomplete for warm 
Jupiters at $\gtrsim 0.3 \AU$. For these systems, a thorough analysis of 
incomplete orbits and trends in RV is required. Unlike close 
solar-type companions, low-mass stellar and brown dwarf companions are unlikely to be excluded from the RV samples to search for 
planets. The combined efforts of RV linear trends and 
high-contrast imaging will yield excellent constraints for such perturbers \citep[e.g.,][]{TREND}.}

Rossiter-Mclaughlin effects for transiting planets are an important diagnostic 
for high-$e$ migration, to which the spin-orbit misalignments have been 
commonly attributed. No ground-based surveys have so far detected 
transiting warm Jupiters 
$(\af = a(1-e^2)>a_{\rm f, crit}=0.1\AU)$.\footnote{Note that 
the Kepler-30 system contains a warm Jupiter and the orbits of its three
planets are shown to be aligned with the spin axis of the host \citep{kepler30}. The three planets are in a compact orbit configuration, and they are unlikely
to be formed by high-$e$ migration.} Yet it is interesting to 
note that, among the ground-based transiting planets with the longest 
period, possibly requiring eccentricity oscillations for tidal migration,
several have known additional planet companions or have large
RV linear trends (e.g., HAT-P-17b \citealt{hatp17b}, 
WASP-8b \citealt{wasp8b}, KELT-6b \citealt{kelt6b}). 
Future ground-based surveys or space-based surveys 
targeting bright stars are likely to discover warm Jupiters suitable for spin-orbit alignment measurements (note a possible transiting warm Jupiter candidate with a
strong perturber identified by \citealt{dawson}). They will be particularly 
interesting candidates subject to our proposed observational test on perturbers.
If the spin-orbit misalignments are solely due to 
high-$e$ migration, and given that the majority of low-$e$ warm
Jupiters do not seem to have strong enough perturbers 
for high-$e$ migration, we expect that the majority of warm 
Jupiters with low-$e$ ($e\lesssim0.2$) will be found to be 
aligned with the spin axes of their hosts.

{Finally, if the warm Jupiters are indeed migrating due to 
tidal dissipation at the high-$e$ stage during Kozai-Lidov
oscillations, they should be tidally powered and 
luminous enough to be detected by the future 
high-contrast imaging facilities such as those to be 
installed at TMT, GMT and ELT \citep{direct}. Similar high-$e$ migration mechanisms have
also been raised for the formation of close binary stars at $P\lesssim 10 \days$
\citep{ft07, binary}, and the constraint we derive in this work can also be applied to test the formation of binaries 
at $10 \days \lesssim P \lesssim 100 \days$ due to high-$e$
mechanisms.}

We thank Andy Gould, Scott Tremaine and Cristobal Petrovich for discussions. 
We are grateful to the referee for a helpful report.
S. D. was partly supported through a Ralph
E. and Doris M. Hansmann Membership at the IAS and 
by NSF grant AST-0807444. B. K. is supported
by NASA through the Einstein Postdoctoral Fellowship
awarded by Chandra X-ray Center, which is operated by
the Smithsonian Astrophysical Observatory for NASA under
contract NAS8-03060. BK and AS acknowledges support from a John N. Bahcall Fellowship at the Institute for
Advanced Study, Princeton. This research has made use of the Exoplanet Orbit Database
and the Exoplanet Data Explorer at exoplanets.org.

\newpage
\begin{figure}
\includegraphics[scale=0.5]{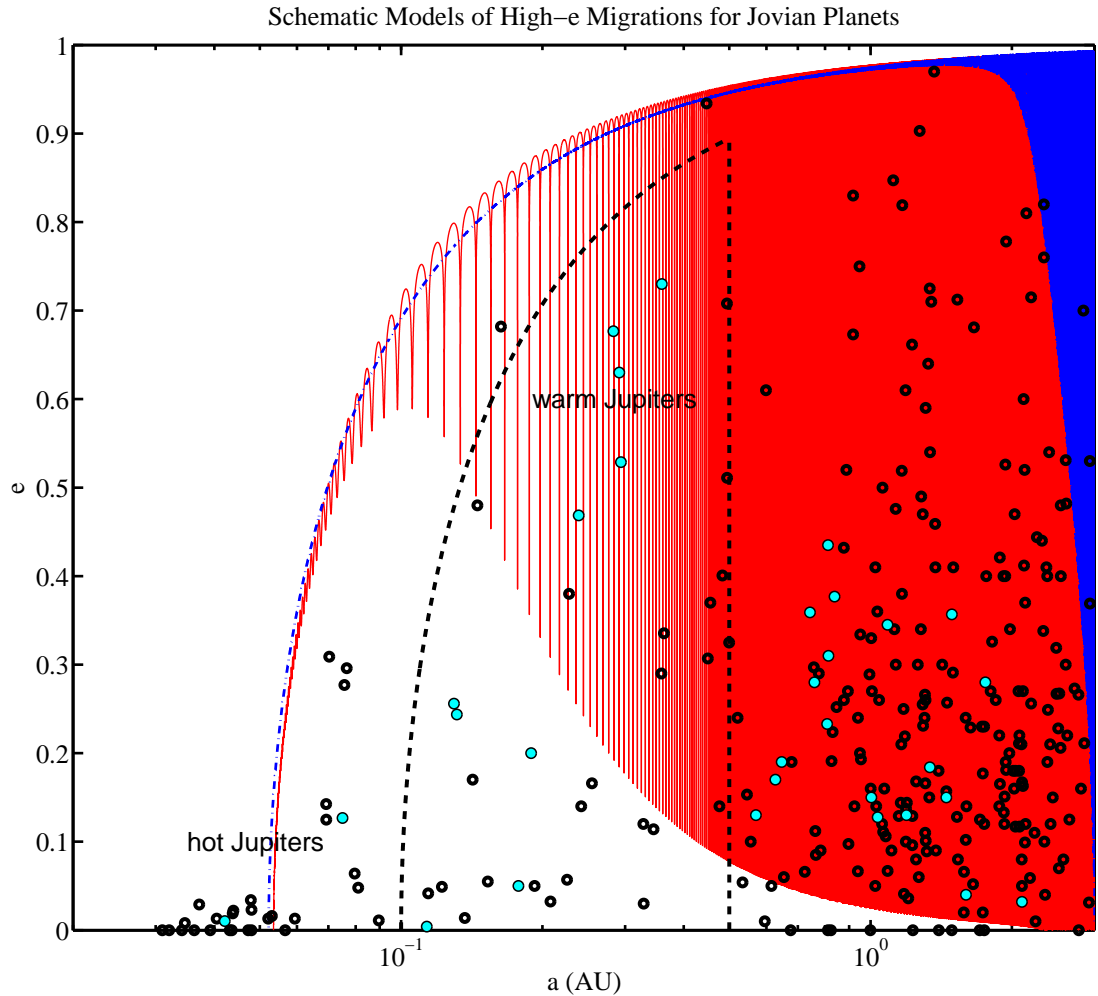}
\caption[width=\textwidth]{Schematic illustrations of high-$e$ migration
on $a-e$ diagram. All Jovian planets discovered by RV are plotted in dots (those with 
additional known Jovian companions in cyan dots while
others in black). 
Warm Jupiters are bounded by the black dashed lines. They are too 
close ($a\lesssim0.5\AU$)  to be formed 
{\it in situ} and too distant to experience efficient
tidal dissipation ($\af = a(1-e^2)>0.1\AU$).
The red solid line shows a possible evolution path to produce a 
warm Jupiter. The gravitational perturber is 
strong enough to overcome GR precession so that the planet 
has significant oscillations in eccentricity at $a \sim 0.3\AU$
 to access $\af\lesssim 0.1\AU$. The observed 
warm Jupiters may be in the low-$e$ stages during such a 
migration from $a\gtrsim 1\AU$. The blue dotted-dashed line shows the high-$e$
migration due to a weak perturber that cannot compete with  
GR near $a \sim 0.3\AU$. Its eccentricity 
does not oscillate and $\af = a(1-e^2)$ is ``frozen'' to a 
low value, $<0.1 \AU$.
\label{fig:schematic}}

\end{figure}
\begin{figure}
\includegraphics*[width=\textwidth]{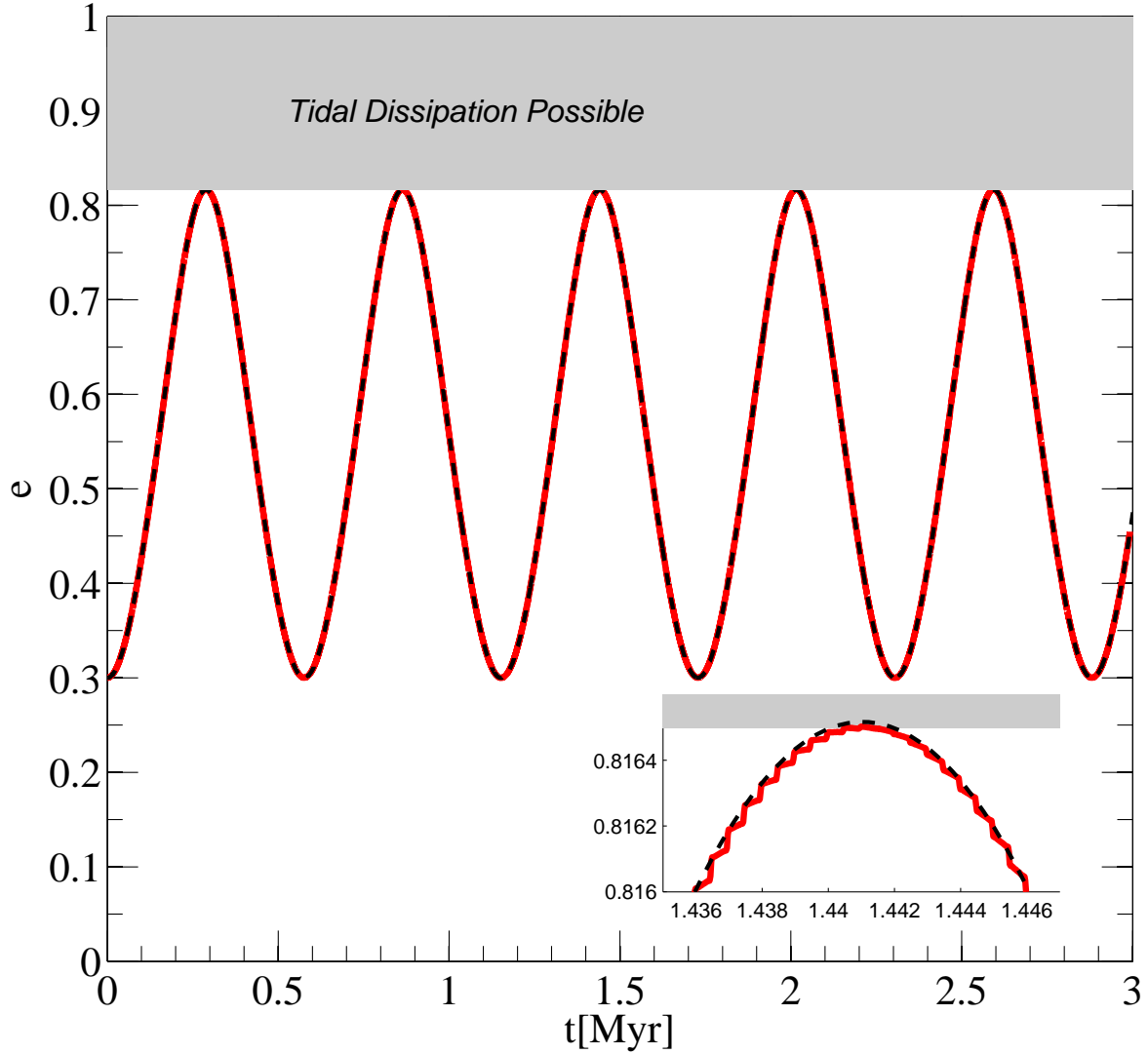}
\caption{
Numerical simulations of Kozai-Lidov oscillations with GR precession
for a warm Jupiter. The planet is at $a=0.3\AU$, $e=0.3$, and
$i=90^{\circ}$, and it has a solar-mass perturber at 
$e_{\per} = 0.5$ and 
$b_{\per} = a_{\per}(1-e_{\per}^2)^{1/2}=68.8 \AU$, which is at the
limit derived from Eq.~\ref{eq:crit} to reach
$\af = a(1-e^2) = a_{\rm f,crit}=0.1\AU$. At higher $e$ (lower $\af$),
the tidal dissipation may be efficient.
The eccentricity as a function of time from direct 3-body
integration is shown in red line and that from double-averaging
calculations (to octupole order in the perturbing potential)
is shown in black dashed line. The two integrations show excellent 
agreement, validating the double-averaging approximation. As can be 
seen in the inset, the 3-body integration shows slight variations 
from the double-averaging calculations within each orbital period 
of the outer perturber. However, their impact on the long 
term evolution ``averages out'' to essentially zero, meaning 
that they play no role in the current study.
}
\end{figure}
\begin{figure}
\includegraphics*[scale=0.6]{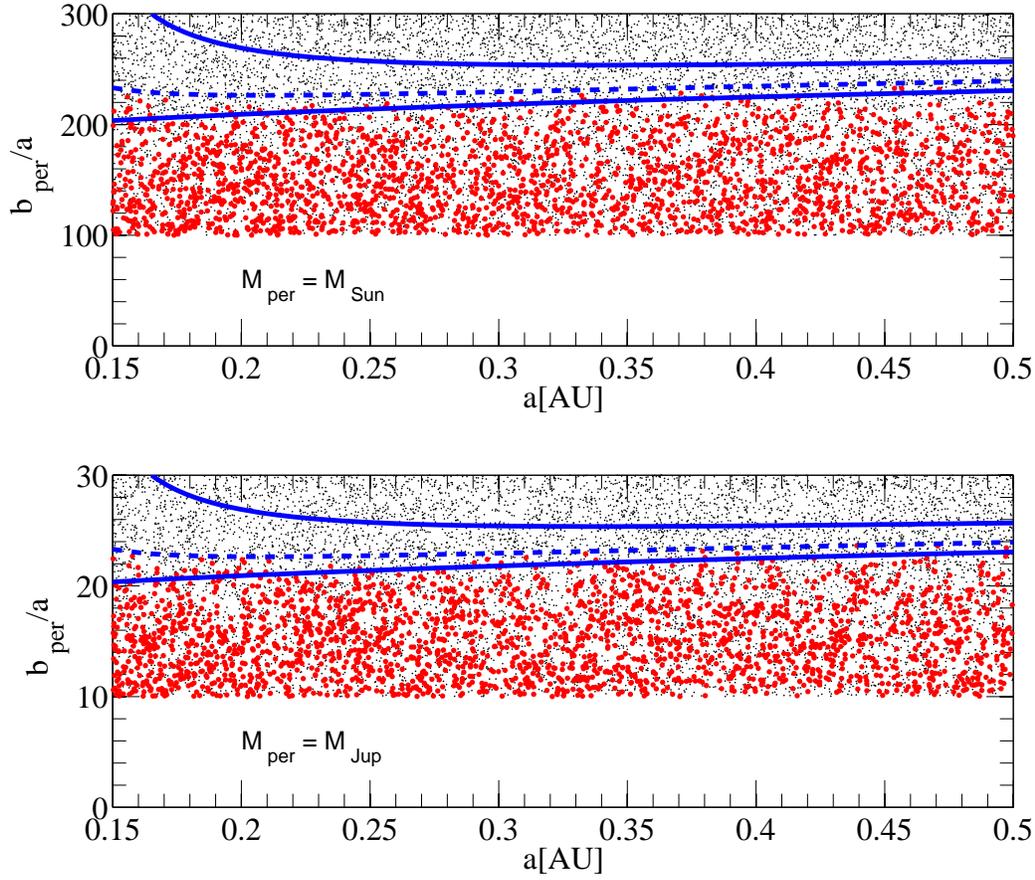}
\caption{The perturber constraint of warm Jupiters
for high-$e$ migration. Upper and lower panels
are for solar-mass and Jupiter-mass perturbers, respectively.
The blue 
lines show the analytical upper limit (Eq.~\ref{eq:crit}) in the 
ratio between semi-minor axis of the perturber and the warm Jupiters'
semi-major axis $b_{\per}/a = (1-e_\per^2)^{1/2} a_\per/a$ as a 
function of $a$. The blue lines from above to below correspond
to eccentricities of warm Jupiters of $e_0 = 0.5, 0.3, 0.0$, 
{which require increasing amount of oscillation amplitude to 
reach the required maximum eccentricity for decreasing 
initial eccentricities}. This 
is verified by 10000 numerical simulations with random 
initial orbital orientations that include the 
double-averaging octupole-order approximation and without 
neglecting the effect of the mass of the warm Jupiter. Each 
simulation is shown as a dot and the initial eccentricity of 
the planet is fixed at $0.3$. Red dots represent 
integrations in which the planet reaches 
$\af = a(1-e^2)<0.1\AU$ within $5{\,\rm Gyr}$ and black otherwise.
The results are in excellent agreement with the corresponding 
analytical constraints shown in blue dashed lines ($e_0=0.3$).
}
\end{figure}
\begin{figure}
\includegraphics*[width=\textwidth]{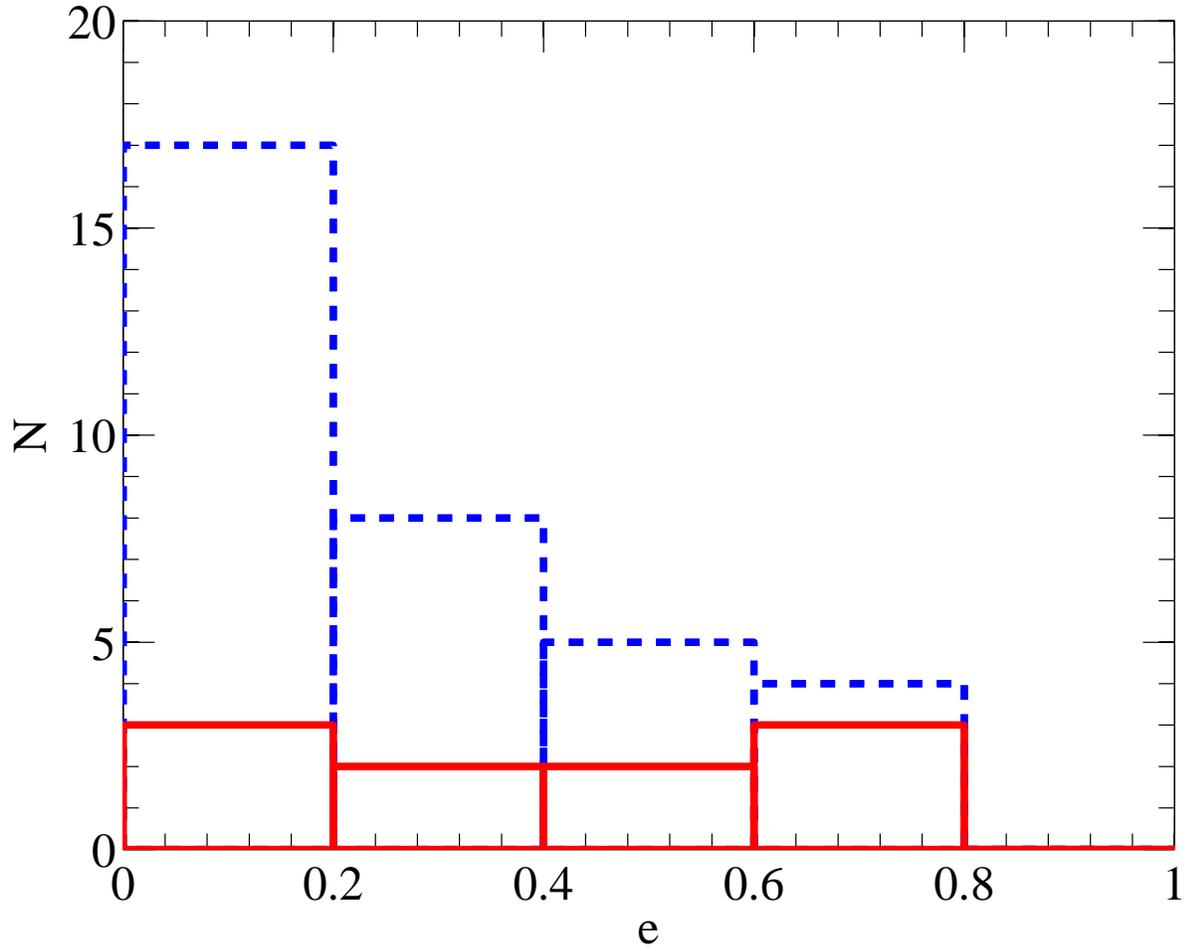}
\caption{The eccentricity distribution of warm Jupiters. 
The blue-dashed histogram is for known RV warm Jupiters 
($M_p \sin i > 0.3 M_{\rm Jup}$, $\af>0.1\AU$, $a<0.5\AU$). 
The red solid histogram is for warm Jupiters with an external Jovian perturber.
All satisfy the constraint in Eq.~\ref{eq:crit}. The fraction of warm Jupiters with detected
Jovian perturbers is a growing function of 
eccentricities. $\gtrsim 50\%$  
of the warm Jupiters with large eccentricities ($e\gtrsim 0.4$) have 
Jovian companions. A large fraction of low-$e$ warm Jupiters 
lack such perturbers. Out of the five warm Jupiter systems with 
$e<0.4$ with known additional Jovian companions, three are in 
compact multiple planet systems with 3 or more planet, which are difficult
to be explained in high-$e$ migration. 
}
\end{figure}
\end{document}